\documentclass[12pt,preprint]{aastex}
\slugcomment{Accepted to ApJ (June 2, 2009)}

\shorttitle{Interferometric observations toward MMS 6}
\shortauthors{Takahashi et al.}

\begin{document}

\title{
Evolutionary Status of Brightest and Youngest Source in the Orion Molecular Cloud -3 Region}

\author{Satoko Takahashi}
\affil{Academia Sinica Institute of Astronomy and Astrophysics, P.O. Box 23-141, Taipei 10617, Taiwan; $satoko{\_}t@asiaa.sinica.edu.tw$ }
\author{Paul T. P. Ho}
\affil{Academia Sinica Institute of Astronomy and Astrophysics, P.O. Box 23-141, Taipei 10617, Taiwan \& Harvard-Smithsonian Center for Astrophysics, 60 Garden Street Cambridge, MA 02138, U.S. A.}
\author{Ya-Wen Tang}
\affil{Department of Physics, National Taiwan University No 1, Sec. 4, Roosevelt Road, Taipei 10617, Taiwan \& Academia Sinica Institute of Astronomy and Astrophysics, P.O. Box 23-141, Taipei 106, Taiwan}
\author{Ryohei Kawabe}
\affil{Nobeyama Radio Observatory, Nobeyama, Minamimaki,Minamisaku, Nagano, 384-1305, Japan \& National Astronomical Observatory of Japan, Osawa 2-21-1, Mitaka, Tokyo, 181-8588, Japan}
\author{\& \\
Masao Saito}
\affil{ALMA Project Office, National Astronomical Observatory of Japan, Osawa 2-21-1, Mitaka, Tokyo 181-8588, Japan}

\newpage

\begin{abstract}
The brightest continuum source in the Orion Molecular Cloud-3 region (OMC-3), MMS 6, was observed 
with the Very Large Array (VLA), the Nobeyama Millimeter Array (NMA), and the Submillimeter Array (SMA). 
Our data were supplemented by near- to mid-infrared archival data taken by Spitzer Space Telescope. 
The compact continuum source, MMS 6-main, was detected with an H$_2$ mass of 3.0 $M_{\odot}$ with a size of 510 AU.
Despite its compact and well condensed appearance, neither clear CO outflow, radio jet, nor infrared sources 
(at a wavelength shorter than 8 $\mu$m) were detected at MMS 6-main even with the present high-spatial resolution 
and high-sensitivity observations. 
The derived H$_2$ column density, 2.6$\times$10${^{25}}$ cm$^{-2}$, corresponds to a visual extinction of $A_v{\sim}$15000 mag., 
and the derived number density is  at least two orders of magnitude higher than for the other OMC-2/3 continuum sources. 
The volume density profile of the source was estimated to have a power-law index of 2 or steeper down to a radius of $\sim$450 AU.
The time scale to form a protostar at the center or the time scale elapsed after its formation is estimated to be 830 to 7.6${\times}10^3$ yr. 
This is much shorter than the typical lifetime of the Class 0/I protostars, which is ${\sim}10^{(4-5)}$ yr, 
suggesting that MMS 6-main is probably in either the earliest stage of the proto-stellar core or in the latest stage of the pre-stellar phase.
\end{abstract}

\section{INTRODUCTION}
Stars form in dense molecular cloud cores ($n{\sim}10^{4-5}$ cm$^{-3}$), which are the precursors of protostars. 
In order to understand the central steps of the star-forming process, it is essential to study the structures, kinematics, and physical conditions of the parental molecular core. These properties provide the initial conditions for star-formation. 

One crucial method to trace the evolutionary status of the parental molecular cores is to investigate 
their density structures. Theoretical studies by Shu et al. (1987) have proposed the collapsing core model 
known as the ``inside-out collapse model''. Initially hydrostatic cores, with the density distribution of $\rho(r){\propto}r^{-2}$ of a singular isothermal sphere, will dynamically collapse from the center with an expansion wave which propagates outward at the sound speed, $C_s$. 
This infalling envelope has the size of $C_{s}t$ at the given time, $t$, and shows a density structure of $\rho(r){\propto}r^{-1.5}$. 
For the alternative semi-analytical isothermal similarity solution for the collapse 
(Larson 1969; Penston 1969), the compression wave propagates inward at the sound speed, reaching the center at $t=0$. 
At $t >0$, this wave is reflected as an expansion wave, and propagates outward with the 
sound speed, $C_s$. The density structure of this collapsing core is the same as for the inside-out collapse case,  although 
the absolute value of the density is higher than the equilibrium case (4.4 times denser for the Larson-Penston solution).

A number of prestellar- and protostellar-cores have been studied using infrared extinction maps and millimeter/submillimeter emission 
images toward low-mass star-forming regions 
(e.g., Ward-Thompson 1994; Ward-Thompson 1999; Andre et al. 1996; Bacmann et al. 2000; Shirley et al. 2000; Kandori et al. 2005; 
Motte et al. 2001; Looney et al. 2000; Harvey et al. 2003$a,b$). 
It has been suggested that the observed density structures of low-mass protostellar cores are roughly consistent with those predicted by 
either the inside-out collapse model or the isothermal sphere model with the density profile $\rho(r){\sim}r^{-1.5}$ or $\rho(r){\sim}r^{-2}$. 
On the other hand, observed prestellar cores show a shallower density power-law index (i.e., flattened density profile) at the core center. 
This can be explained if (i) the core is not sufficiently evolved to have reached the isothermal sphere density profile (i.e., t$<$0 case in the Larson-Penston case), 
(ii) other mechanism such as a magnetic field could support the system (e.g., Crutcher et al. 1994), or (iii) the core has a density structure 
explained by a finite-sized Bonner-Ebert sphere (e.g., Bonnor 1956).  
A large sample of these studies is one of the crucial strategies for understanding the initial conditions of the star-forming core 
as well as investigating the evolutionary status of the central stars. 
In particular, high-resolution observational studies with a spatial resolution of much less than 1000 AU are necessary to constrain the radius 
where the density profile steepens and to compare this with the theoretical model for the very early stages of protostellare cores. 
Such higher angular resolution studies are still limited in number.

The OMC-2/3 region is located at the northern-part of the Orion Molecular Cloud A ($d{\sim}400$ pc; Sandstrom et al. 2007; 
Menten et al. 2007; Hirota et al. 2007; Kim et al. 2008), which is the nearest giant molecular cloud, and one of the best-studied cluster-forming regions at all 
observable wavelengths (e.g., Bally et al. 1987, Allen et al 2007; Tatematsu et al. 1993, Johnstone \& Balley. 1999; Tsujimoto et al. 2005). 
In this region, filamentary-shaped molecular clouds, lying in the north-south direction 
with several pc scale, were discovered in the submillimeter continuum observations with the IRAM 30 m telescope and the JCMT/SCUBA 
(Chini et al. 1997; Johnstone \& Bally. 1999; Johnstone et al. 2003). These filamentary structures contain a few tens of 
millimeter and sub-millimeter continuum sources, presumably tracing prestellar- and protostellar-cores 
(Chini et al. 1997; Lis et al. 1998; Nielbock et al. 2003; Takahashi et al. 2006; Takahashi 2008$a,b$; Shimajiri et al. 2008). 
In addition, approximately a dozen radio jets, molecular outflows, and shock excited H$_2$ emissions, were detected in this region 
(Yu et al. 1997; Reipurth et al. 1999; Aso et al. 2000; Stanke et al. 2002; Williams et al. 2003; Matthews et al. 2005). 
Moreover, our recent extensive molecular outflow survey with ASTE in the CO(3--2) emission have revealed 14 molecular 
outflows, which includes seven newly detected outflows, suggesting that the OMC-2/3 region is one of the most active nearby on-going 
star-forming regions and contains the richest assemblages of protostars (Takahashi et al. 2008$b$).

From a series of our OMC-2/3 survey observations combined with previous millimeter and submillimeter continuum studies, 
we have found an interesting 1.3 mm continuum source, MMS 6, which is the brightest source in the millimeter- and submilimeter-wavelengths 
in the OMC-2/3 region.  
This brightest source is located at the center of the OMC-3 region, and the 1.3 mm flux is roughly one order of magnitude larger than 
those for any other continuum sources in OMC-2/3 (Chini et al. 1997; Johnstone \& Bally 1999). 
Despite the unusual appearance of MMS 6, no signature of star-formation activities such as molecular outflow or jet have been detected toward MMS 6 
by Takahashi et al. (2008$b$). The bolometric luminosity, temperature, and core mass of MMS 6, which were derived from the previous single-dish 
millimeter to submillimeter observations combined with IRAS data, are $<$60$L_{\odot}$, 15-25 K, and 36 $M_{\odot}$, respectively (Chini et al. 1997). 
The derived luminosity and mass are one order of magnitude higher than the values derived in typical low-mass star-forming 
regions such as the Taurus molecular cloud (e.g., Myers \& Benson 1983). 

That MMS 6 does not show any significant signatures of star-formation activities, suggests that MMS 6 is at the earliest stage of evolution. 
To address this issue, our primary goal is to constrain the physical conditions such as size, mass, density, density profile, and dust grain properties. 
Comparison of the derived values with the low-mass core values is also important to elucidate the star-formation environment. 
There is another close-by 1.3 mm source, MMS 5, which is located at 25$''$ north-west of MMS 6. 
We have detected a clear bipolar outflow associated with MMS 5 in the CO(3--2) emission, which is mainly elongated in the east-west direction. 
While there may be additional diffuse gas extending to the MMS 6 direction, the relatively large beam of the previous CO (3--2) observations (26$''$) makes 
it difficult to investigate the detailed spatial structures of molecular outflows and the physical properties of cores in the MMS 5/6 regions. 

To unveil the evolutionary status of MMS 6, in this paper, we have investigated the detailed physical properties 
with higher spatial resolution observations using the VLA, the NMA, and the SMA. 
Moreover, near- to mid-infrared data retrieved from the Spitzer Space Telescope (IRAC and MIPS) archival data, 
and near-infrared JHK$_s$ data taken with the SIRIUS/IRSF (taken by Takahashi et al. 2008$b$), were compared with the 
interferometric data with ${\sim}1''$ relative position accuracy. 

The details of the millimeter- and submillimeter-interferometric observations are presented in Section 2. 
Results of the millimeter- to infrared-continuum emissions and also some molecular lines are shown in Section 3. 
Physical properties of MMS 6 are discussed in Section 4. Finally, Section 5 summarizes the paper and shows future prospects.

\section{OBSERVATIONS AND DATA REDUCTION}

\subsection{Interferometric Observations}	
	Aperture synthesis observations of molecular lines and continuum emissions were carried out 
	using the Nobeyama Millimeter Array (NMA), the Submillimeter Array (SMA), and the Very Large Array (VLA). 
	The phase reference center was set on the peak position of the 1.3 mm source, 
	MMS 6, which has a coordinate of R.A. (J2000) = 5$^h$35$^m$23.48$^s$, 
	Decl.(J2000) = -05$^{\circ}$01$'$32$''$.20 (Chini et al. 1997). 
	The observed parameters for the continuum and molecular line observations are summarized in Table 1 and Table 2, respectively.
	
	\subsubsection{NMA Observations}
		The 3.3 mm and 2.3 mm continuum emission were observed with the six 10 m antennas of the Nobeyama Millimeter Array (NMA) 
		\footnote{Based on the observations made at the Nobeyama Radio Observatory, which is a branch of the National Astronomical Observatory, 
		an inter university research institute operated by the Ministry of Education, Culture, Sports, Science, and Technology of Japan}. 
		H$^{13}$CO$^{+}$ ($J$= 1--0; 86.754 GHz) emission was observed together with the 3.3 mm continuum emission, and 
		$^{12}$CO ($J$= 1--0; 115.271 GHz) emission was observed independently. 
		These observations were performed during May 2004 to January, 2006. The 3.3 mm and 2.3 mm continuum data were obtained 
		using a digital wideband correlator, UWBC, which has a bandwidth of 1024 MHz and 512 frequency channels for each of the lower- and 
		upper-sidebands (LSB and USB; Okumura et al. 2000). 
		The LSB and USB continuum data separated by 12 GHz in RF were combined to improve the sensitivity. 		
		Spectral information of these lines were obtained with the 32 MHz bandwidth FX correlator containing 1024 channels. 
		The rest frequencies of the H$^{13}$CO$^{+}$ and $^{12}$CO lines were placed in the center of the LSB and USB, respectively. 
		The phase and amplitude calibrator, 0531+135, was observed every 20 minutes.
		Observations of Uranus provided the absolute scale for the flux density calibration.	
		The overall flux uncertainty was estimated to be $\sim$15\%.
		The passband across the bandwidth was determined from observations of 3C 454.3 at 3.3 mm and 0530+135 at 2.3 mm 
		with a 30 minute integration. The combined arrays provided projected baselines ranging from 3 to 115 k$\lambda$ 
		at 3.3 mm and 4 to 38 k$\lambda$ at 2.3 mm.
		
		The raw data were calibrated for the complex gain and passband using the standard reduction program, 
		UVPROC-II (Tsutsumi, Morita, \& Umeyama 1997). 
		After the calibrations, final CLEANed images were made using the Astronomical Image Processing System (AIPS), 
		developed at NRAO\footnote{NRAO is a facility of the National Science Foundation, operated under cooperative agreement by Associated 
		Universities, Inc.}. Natural weighting and robust weighting (0 = the weighting between natural and uniform) 
		were applied for the line data and continuum data, respectively.
		The resulting synthesized beam sizes, the achieved rms noise levels as well as other observational parameters are 
		summarized in Table 1 and  Table 2 for the NMA molecular lines and continuum observations , respectively.

	\subsubsection{SMA Observations}
		 The 0.9 mm continuum data were taken using the eight 6 m antennas of the Submillimeter Array
		 \footnote{The Submillimeter Array is a joint project between the Smithsonian Astrophysical Observatory and the 
		 Academia Sinica Institute of Astronomy and Astrophysics and is funded by the Smithsonian Institution and the Academia Snica.} 
		 (SMA; Ho et al. 2004), on January 2 and October 21, 2008. 
		 The combined arrays provided projected baselines ranging from 13 to 88 k$\lambda$.
		 Both the LSB and USB data were obtained simultaneously with the 90$^{\circ}$ phase switching technique by the 
		 digital spectral correlater, which has a bandwidth of 2 GHz in each sideband. 
		 The LSB and USB continuum data separated by 10 GHz RF were combined to improve the sensitivity. 
		 The phase and amplitude calibrator, 0531+135 (1.7 Jy), was observed every 20 minutes. 
		 Observations of Titan provided the absolute scale for the flux density calibration.	
		 The overall flux uncertainty was estimated to be $\sim$20\%. 
		 The passband across the bandwidth was determined from observations of 3C 273 and 3C 454.3 with a 60 minute integration in each.

		 The raw data were calibrated using MIR, originally developed for the Owens Valley Radio Observatory (Scoville et al. 1993) 
		 and adopted for the SMA. 
		 After the calibration, final CLEANed images were made using the AIPS task IMAGER with a robust weighting (=0). 
		 The resulting synthesized beam size was 2$''$.2$\times$2$''$.0 (or 880 AU $\times$ 800 AU at the adopted distance of 400 pc) 
		 with a position angle of -62$^{\circ}$. The achieved rms noise level was 43 mJy. The observational parameters are summarized in Table 1.

	\subsubsection{VLA Observations}	
		The 3.6 cm data were taken with the Very Large Array (VLA) of NRAO in the CnD-array on June 1, 2008. 
		Both the LSB and USB data were obtained simultaneously with the 90$^{\circ}$ phase switching technique by the 
		digital spectral correlater, which has a bandwidth of 50 MHz in each sideband (total bandwidth of 100 MHz). 
		The array provided projected baselines ranging from 8.7 to 55 k$\lambda$. 
		The phase and amplitude calibrator, 05416-05418 (1.4 Jy), was observed. 
		The 7.3 mm data were also taken with the VLA in the CnD- array. In addition, C-array data retrieved from the NRAO archival 
		system were combined.  
		The combined arrays provided projected baselines ranging from approximately 
		4.5 to 477 k$\lambda$. The phase and amplitude calibrator, 05416-05418 (0.7 Jy), was observed.
		In order to track the short-term amplitude and phase variations induced by the earth atmosphere, 
		the fast switching technique was adopted for both 3.6 cm and 7.3 mm observations. 
		The cycle time of the observing target, MMS 6, and the phase-amplitude calibrator, 05416-05418 which lies about 
		2 $^{\circ}$ from MMS 6, was set at 4.3 minutes. 
		For the 7.3 mm observations, the pointing accuracy of each antenna was checked every 45 minutes from scans of the quasar 0541-056. 
		Absolute flux calibrations at 3.6 cm and 7.3 mm were performed by observations of 3C 286  for the C-array data and 3C48 for the CnD-array data.
		The overall flux uncertainties of 3.6 cm and 7.3 mm observations were estimated to be a few percents. 

		The raw data were calibrated using the AIPS software package. 
		After the calibrations, we made final images using the AIPS task IMAGR with robust weighting (=0). 
		The resulting synthesized beam sizes at 3.6 cm and 7.3 mm were  
		10$''$.0$\times$3$''$.0 (or 4000 AU $\times$ 1200 AU at an adopted distance of 400 pc) with a position angle of 67$^{\circ}$ 
		and $1''.4{\times}0''.6$ (or 560 AU $\times$ 240 AU) with a position angle of 
		-87$^{\circ}$, respectively. The achieved rms noise levels at 3.6 cm and 7.3 mm were 50-60 $\mu$Jy and 19 mJy, respectively. 	 
		Further 3.6 cm and 7.3 mm observational parameters are summarized in Table 1.

\section{RESULTS} 
\subsection{Source Identification Based on Continuum Images}
	\subsubsection{Millimeter and Submillimeter Continuum Emission}
			Figure 1 shows the four-band continuum maps (i.e., 0.9, 2.3, 3.3, and 7.3 mm) toward MMS 6 taken with the VLA, NMA, and SMA. 
			The single continuum source detected by the single-dish 1.3 mm, 850 $\mu$m, and 350 $\mu$m observations (Figure 1$a$), 
			MMS 6, was resolved to be a strong main 
			component (hereafter MMS 6-main) with additional substructure by the interferometric observations. 
			To measure the size, position, and flux, 2D-Gaussian fitting was applied to each map, and the derived parameters are summarized in Table 3. 
			
			MMS 6-main, which has the peak position of [RA(J2000)=5$^h$35$^m$23.43$^s$, Dec(J2000)=-5$^{\circ}$01$'$30.6$''$], 
			was detected at all the observed millimeter- and submillimeter- wavelengths. The peak position of the 7.3, 3.3, 2.3, and 0.9 mm 
			data coincides within the positional accuracies of these data, $\leq$0.5$''$. These peak positions are located at 
			approximately 2$''$ north-west of the 1.3 mm continuum emission peak, measured with the IRAM 30 m telescope (Chini et al. 1997), 
			which has a positional accuracy of ${\pm}5''$. 
			The flux density of MMS 6-main measured with the SMA 0.9 mm continuum observations is 4.0 Jy, corresponding to 
			60 \% of the flux measured with the 13 $''$ beam 
			of the JCMT/SCUBA 850 $\mu$m observations, which is 6.6 Jy (Johnstone \& Balley. 1999). 
			
			A second component, which is located at ${\sim}1.5''$ north-east of the MMS 6-main (hereafter MMS 6-NE), 
			was clearly detected at 0.9 mm. The other three-wavelength data also show a hint of the additional component with 3 to 6 $\sigma$ level emission. 
			2D-Gaussian fitting was applied to this second component in the 0.9 mm data. 
			The peak position at the 0.9 mm emission coincides with the previously detected 3.6 cm source position by Reipurth et al. (1999) 
			with a positional offset of 1.0$''$, which is comparable to the absolute positional accuracy of the 3.6 cm observations (${\sim}1''$).
			
			Extended emission from the large-scale filamentary structure, which was detected in the millimeter- and submillimeter- single-dish 
			observations (Figure 1$a$), was not detected in the interferometric observations. 
			As described in Table 1, the maximum detectable angular sizes for the interferometric observations, 
			were 13$''$--56$''$. Therefore, the structures, which have a size scale larger than the maximum detectable sizes were missed in our images. 
			
			In addition, another 1.3 mm source, MMS 5 identified by IRAM 30 m observations (Chini et al. 1997) was detected 
			at approximately 22$''$ north-west of MMS 6-main with the 7.3 mm and 3.3 mm continuum emissions with 3--6 $\sigma$ 
			level signals (Figure 1$c$ and $d$).

			\subsubsection{Near- to Mid- Infrared Sources}
			Figure 2$a,b$ show the 2.2 $\mu$m and 8 $\mu$m images overlaid with the 3.3 mm continuum emission. 
			From the images, a total of five infrared sources located on the main OMC filament detected at either 2.2 $\mu$m-, or 8 $\mu$m-bands, or both 
			(i.e., denoted by Figure 2 doted circles) were identified within $\sim$0.5 pc radius from MMS 6-main (see Table 4). 
			From the SEDs and color-color-diagrams, three of the sources (IRS 1, IRS 3, and IRS 5) have been identified as protostars 
			(Peterson \& Megeath 2008; see also Table 4). 
			
			Of the five detected infrared sources, IRS 3 is the only one closely associated with the strong continuum source, MMS 6. 
			The position of IRS 3 was measured in the Spitzer/IRAC images
			\footnote{We found that the peak position of IRS 3 as measured in the $Ks$ image is roughly 1.5$''$ shifted to the west 
			from the peak position as determined from  Spitzer/IRAC band images, which all self consistent to within $\sim$0.5$''$. 
			We have checked the positional alignments of the other point-like source in the same frame between different data sets 
			(i.e., comparison of the SIRIUS/IRSF 
			and Spitzer/IRAC data based on the 2MASS data), 
			and found the point-like sources in the same frame coincide with each other. 
			Hence the IRS 3 positional offset between the Ks-band and Spitzer/IRAC images is considered to be real.
			The inward extinction of $K_s$ most likely does not allow the true position of IRS 3 to be detect at $K_s$}.
			MMS 6-main and MMS 6-NE, are located at $2''.1$ and $0''.8$ away from IRS 3. 
			The important question is whether IRS 3 is a 
			heating source of the millimeter/submillimeter sources, MMS 6-main or MMS 6-NE, or not?
			A comparison of the relative positions between the infrared data and the interferometric data in the context of the accuracy of 
			the positional alignment between the images is necessary.
			The positional alignment of the near- to mid-infrared images, taken by SIRIUS/IRSF and Spitzer/IRAC, is determined 
			by the standard stars in the observed frames, and the positional accuracy of these data is determined by the positional alignment accuracy 
			of the 2MASS point source catalog. In this case, both near- and mid-infrared images have the same positional accuracies with 
			a typical value of $<{\pm}1''.2$.  
			The relative positions of the interferometric data are determined by the positional accuracy of the gain calibrators, 
			which typically is $<{\pm}0''.5$. The best way to align the infrared and interferometric images is to match the common sources in 
			the observed field. 
			However, no common standard star was identified within our observed fields of view.
			Instead, we use another nearby OMC-3 millimeter source, MMS 7, observed by Takahashi et al. (2006), 
			which was detected in both the interferometric 3.3 mm data taken with the NMA and the 8 $\mu$m data taken with Spitzer/IRAC.
			In the 8 $\mu$m data, both MMS 7 and MMS 5/6 are in the same processed image. The 3.3 mm observations of MMS 7 and 
			MMS 5/6, while not in the same primary beam, were made during the same 2004 November to 2005 March  period with 
			same array configurations. 
			Since the positions of MMS 7 at 3.3 mm and 8 $\mu$m coincide to within 0.31$''$, 
			the accuracy of the positional alignment between the infrared- and interferometric images is on this order. 
			The positional offset between MMS 6-main and IRS 3, at 2.1$''$, is formally a 7$\sigma$ result. 
			This suggests that the infrared source, IRS 3, is likely not associated with MMS 6-main. 
			On the other hand, the peak position of MMS 6-NE which was detected in the 0.9 mm continuum image, 
			coincides with IRS 3 within the positional accuracies. This suggests that IRS 3 may be associated with the dusty envelope 
			detected in the 0.9 mm emission. 
			
			Another infrared source, IRS 1, is located at the position of the 1.3 mm continuum source, MMS 5 which also has been detected 
			in the interferometric observations at 3.3 mm and 7.3 mm. The infrared source was only detected in the wavelength longer than 
			the Spitzer/IRAC 5.6 $\mu$m-band, and the source has a Class 0 type SED.

\subsection{Molecular Line Emissions}
	\subsubsection{Search for high-velocity gas components in the MMS 5/6 region; CO(1-0) Emission}
			Figure 3 shows the blueshifted ($V_{\rm{LSR}}$=8.3 -- 10.3 km s$^{-1}$) 
			and redshifted ($V_{\rm{LSR}}$=12.5 -- 14.5 km s$^{-1}$) ranges of the $^{12}$CO(1--0) emissions towards the MMS 5 
			and MMS 6 region, superposed on the $Ks$-band image taken with the SIRIUS/IRSF by Takahashi et al. (2008$b$). 
			
			A pair of blueshifted and redshifted components are seen in the east-west direction at $\sim$7$''$ east of MMS 6-main 
			with a velocity less than 3.5 km s$^{-1}$ from the systemic velocity. 
			Distribution of the redshifted emission is highly consistent with the jet-like feature detected in the $K_s$-band image emanating from IRS 4. 
			The extended blueshifted component is elongated along the northeast to southwest. 
			The counterpart  of the blueshifted component is not detected in the $K_s$-band. 
	
			The extended redshifted emission with a velocity of 14.3 km s$^{-1}{\leq}{V_{\rm{LSR}}}{\leq}$ 15.4 km s$^{-1}$ 
			was also detected in the CO(3--2) emission by the single-dish ASTE 10 m telescope (Takahashi et al. 2008$b$) with the same position as 
			the CO(1--0) emission detected with the NMA.  At the position of MMS 6-main, no clear signature of typical molecular outflows, 
			which has highly collimated bipolar structure with high-velocity component, was detected in the interferometric CO(1--0) 
			observations with these high-spatial resolution (i.e., $\sim$3000 AU) 
			and high-sensitivity (i.e., $\sim$ 6.5$\times$10$^{-4}~M_{\odot}$ ) observations.
	
			Another red-shifted component was detected to the east of MMS 5. 
			The redshifted emission in the velocity range of $V_{\rm{LSR}}$ = 13.2-- 21.9 km$^{-1}$ was 
			already detected in the CO(3--2) emission with the ASTE 10 m telescope (Takahashi et al. 2008$b$).  
			The distribution of the redshifted component detected by the NMA is consistent with that found by the ASTE CO(3--2) observations.  
			The blue-shifted component associated with MMS 5, which was detected in the ASTE CO(3--2) observations, 
			was not detected in the CO(1-0) emission,  
			probably due to the primary beam attenuation of the NMA.

	\subsubsection{Dense Gas Distribution; H$^{13}$CO$^{+}$ (1--0) Emission}
			In Figure 4, H$^{13}$CO$^{+}$ shows an elongated structure from northwest to southeast with a ridge-like structure from 
			the MMS 5 to MMS 6 region. This elongation agrees well with the elongation of the large-scale filamentary structure detected by single-dish 
			observations at 1.3 mm, 350 $\mu$m, and 850 $\mu$m wavelengths (Chini et al, 1997; Lis et al. 1998; Johnstone \& Balley. 1999). 
			Assuming that the H$^{13}$CO$^{+}$ emission is optically thin and that the distribution of the excitation temperature is uniform, 
			the total LTE mass of the entire H$^{13}$CO$^{+}$ emission structure is estimated to be 62--153 $M_{\odot}$ with an average 
			column density of (0.7--1.0)${\times}10^{24}$ cm$^{-2}$. 
			Here, the excitation temperature of 20 K (e.g., Cesaroni et al. 1994, Chini et al. 1997) 
			and the H$^{13}$CO$^{+}$ abundance of (0.45--1.4)${\times}10^{-10}$ (Aso et al. 2000, Takahashi et al. 2006) are adopted. 			
			
			The position of H$^{13}$CO$^{+}$ emission associated with MMS 6-main is established with a 2D-Gaussian fit with a deconvolved size of 
			11$''{\times}7''$.1 (P.A. 9.7$^{\circ}$), which corresponds to 4400$\times$2800 AU at $d$=400 pc. 
			The LTE mass derived from this emission is estimated to be 7.9--25 $M_{\odot}$ with an average hydrogen column density of 
			(0.9--2.7)${\times}10^{24}$ cm$^{-2}$, under the assumptions described in the previous paragraph.
			The peak position of the H$^{13}$CO$^{+}$ emission has a slight offset 
			(${\Delta}$offset=2$''$.5) from the dust continuum peak (i.e., MMS 6-main).
			Although the physical conditions cannot be tightly constrained with only the H$^{13}$CO$^{+}$ (1--0) observations, 
			the peak position offset between the H$^{13}$CO$^{+}$ emission and the continuum emission 
			could be explained by an opacity effect, density differences along the line of sight, 
			or H$^{13}$CO$^{+}$ depletion in the densest region , as traced by the millimeter- and submillimeter- continuum emissions.
		
			The H$^{13}$CO$^{+}$ emission was also detected at MMS 5 with a deconvolved size of 16$''$.0$\times$6$''$.2 (P.A. 180$^{\circ}$), 
			which corresponds to 6400$\times$2500 AU at $d$=400 pc. 
			The LTE mass derived from the emission was estimated to be 7.4--23 $M_{\odot}$ with an average column density of 
			(0.7--2.2)$\times$10$^{24}$ cm$^{-2}$ under the assumptions described in the first paragraph. 
			The H$^{13}$CO$^{+}$ peak coincides with an infrared source IRS 1 and also dust continuum emission 
			detected by our interferometric observations, suggesting that IRS 1 (MMS 5) is still surrounded by the thousands AU scale dense gas envelope.
			No strongly concentrated H$^{13}$CO$^{+}$ emission was detected at the other three infrared positions of IRS 2, 4, and 5. 
			The extended H$^{13}$CO$^{+}$ emission with approximately 15 $\sigma$ level emission was detected at IRS 3, 
			although it does not coincide with any of the H$^{13}$CO$^{+}$ emission peaks. 
			IRS 3 probably is still surrounded by the dense envelope, which was detected in the H$^{13}$CO$^{+}$ emission.

\subsection{Spectral Energy Distribution}			
			Figure 5 shows the flux density of MMS 6-main as a function of the frequency. 
			In order to enhance the sensitivity to the compact structure associated with MMS 6-main, 
			the flux densities were measured in images where the $uv$-range is restricted to be lager than 30 k$\lambda$.  
			We used the natural weighting map to be more sensitive to the extended structure at 7.3 mm since the 
			7.3 mm continuum image shows a smaller size-scale structure. 
			Flux density and deconvolved size of the MMS 6-main compact component at each wavelength are summarized in Table 5.  
						
			3.6 cm (i.e., 8.4 GHz) continuum observations were made by Reipurth et al. (1999) with the VLA-D array toward the MMS 6 region. 
			They have reported the detection of the 3.6 cm continuum emission with a flux density of 0.15 mJy. 
			However, the peak position of the 3.6 cm continuum emission is 2.0$''$ offset from the the peak position of MMS 6-main. 
			This positional offset is larger than the positional accuracy of the 3.6 cm observations (${\sim}1''$), suggesting 
			that the 3.6 cm continuum emission is probably not associated with MMS 6-main. 
			In addition, no 3.6 cm continuum emission was detected in our VLA-DnC array observations with the upper limit (3 $\sigma$ level) of 0.09 mJy.  
			Because of these uncertainties, in this paper, the 3.6 cm flux density reported by Reipurth et al. (1999) is adopted as an upper limit. 
			
			The best fit of the spectral index $\alpha$ (i.e., $F_{\nu}{\propto}{\nu}^{\alpha}$) in Figure 5 is derived to be 2.93. 
			Since the spectral index, $\alpha$, and the dust emissivity, $\beta$, have a 
			relation of ${\alpha}={\beta}$+2,  the dust emissivity, $\beta$, is estimated to be 0.93.
			3.6 cm and 7.3 mm continuum emission sometimes trace an ionized jet (i.e., free-free emission) driven by low- to 
			intermediate-mass protostars (e.g., Anglada 1995; Rodriguez 1997). Hence, if excess emission at centimeter wavelengths 
			(i.e., mainly $\lambda{>}$7 mm for the low- to inter-mediate mass case) were detected, it may indicate the presence of a 
			central heating source. However, no clear excess emission caused by the free-free emission is detected toward 
			MMS 6-main (Figure 5). 
			
			On the assumption that the dust emission is optically thin at 0.9 mm and the temperature distribution of the dust continuum 
			emission is uniform, the mass of the dusty condensation is estimated to be 3.0 $M_{\odot}$ with a size of 
			400$\times$320 AU (see flux and deconvolved size at 0.9 mm in Table 5), and the molecular hydrogen column density is estimated to be 
			2.6${\times}10^{25}$ cm$^{-2}$. 
			Here, we assume that $\beta$=0.93, $T_d$=20 K, and ${\kappa}_{\lambda}$=0.037 cm$^{2} 
			{\rm{g}}^{-1}$ ($\lambda$/400 $\mu$m)$^{-{\beta}}$ (Cesaroni et al. 1994; Chini et al. 1997; Keene et al. 1982). 
			The hydrogen column density and the visual extinction have a relation of $A_v=0.56 N_H +0.23$, where $N_H$ is the equivalent 
			column density in units of 10$^{21}$ cm$^{-2}$ (Predehl \& Schmitt 1995). The visual extinction is estimated from the hydrogen column density 
			to be approximately 15000 mag. Moreover, this visual extinction corresponds to $A_{8{\mu}m}$=790 mag., which is 
			much higher than the value toward YSOs in the OMC filament, from several to 12 mag. (Peterson \& Megath et al. 2008).
			Therefore, it is most likely that the central protostar associated with MMS 6-main cannot be detected because of the extremely 
			high-extinction toward MMS 6-main. 
	
			Similarly, H$_2$ mass of MMS 5 were derived as 0.3 $M_{\odot}$ from 3.3 mm continuum flux 
			with assumptions that $\beta$=0.8, $T_d$=20 K, and ${\kappa}_{\lambda}$=0.037 cm$^{2}$ g$^{-1}$ ($\lambda$/400 $\mu$m)$^{-{\beta}}$. 
			Here, $\beta$ were derived using 3.3 mm and 7.3 mm continuum data. 
			The visual extinction was estimated to be $\sim$220 mag., which corresponds to $A_{8{\mu}m}$=12 mag. 
			This value is comparable to detected YSOs in the OMC filaments (Peterson \& Megath et al. 2008).  
			The central infrared source, IRS 1, was actually detected in the 24 $\mu$m Spitzer/MIPS image.

\section{Discussion: Nature of  Dusty Core: MMS 6-main}
     			We are studying the physical properties of MMS 6-main in order to address its evolutionary status, as an example of 
			the star-forming environment of the OMC-2/3 region. In this section, we discuss the physical properties of 
			MMS 6-main from three points of view; (i) dust grain properties, (ii) comparison of physical properties between MMS 6-main 
			and the other pre- and proto-stellar cores in the OMC-2/3 region, and (iii) internal structure of the cores.
			
			\subsection{Dust Grain Properties}
			The dust emissivity spectral index, $\beta$, is related to dust properties such as grain sizes and grain shapes 
			(e.g., Ossenkopf \& Henning 1994; Pollack et al. 1994; Draine 2006). 
			Observations have shown that $\beta$=1.5--2 in the general interstellar medium or dense molecular clouds/cores 
			(e.g., Goldsmith et al. 1997), and $\beta{\sim}$1 
			in circumstellar disks (e.g., Beckwith \& Sargent 1991; Andrews \& Williams 2007). 

			 The anomalously low dust emissivity index, $\beta$=0.3, in the MMS 6 region were reported from previous 
			 JCMT/SCUBA observations by Johnstone \& Bally (1999). 
			 A similarly low value, $\beta$=0.2, was also derived in the Orion KL region, which is the most active star-forming region in the OMC filament. 
			Johnstone \& Bally (1999) found that only these two regions show such low-$\beta$ values, while the other eighteen bright sources in the 
			OMC filament show an average of $\beta$=1.2. In Section 3.3, the dust emissivity spectral index of MMS 6-main 
			was estimated to be $\beta$ =0.93 from the multi-wavelength interferometric data. 
			The estimated $\beta$ is higher than the value derived from low-spatial resolution JCMT/SCUBA data toward MMS 6, 
			suggesting that there is a distribution in $\beta$ for different size-scales. On the other hand, the derived $\beta$ shows a lower value 
			as compared with the $\beta$ derived from the other bright continuum sources in Orion. 

			In order to consider the behavior of $\beta$ , the spatial distribution of the dust opacity spectral index, $\beta$, 
			is calculated using the 3.3 mm and 0.9 mm continuum images with the following equation,
						
			\begin{equation}
					{\beta} =  \frac{log (F{({\nu}_1)}/F{({\nu}_0}))}{log({\nu}_1/{\nu}_0)} -2	
			\end{equation}
			
			where $F_{\nu}$ is the flux in each pixel, and ${\nu}_1$ and ${\nu}_0$ are the frequencies corresponding to 0.9 mm and 3.3 mm, respectively. 
			For the calculation, only regions which have a signal-to-noise ratio above 3 $\sigma$ were used. 
			Figure 6 presents the spatial distribution of $\beta$ in the MMS 6 region. 
			In Figure 6, there is a suggestion that the $\beta$ value has significantly changed along the north-south direction with 
			a range of $\beta$=0--1.5. Especially, $\beta$ =0--1.5 is suggested at MMS 6-main, while $\beta{\sim}$1.5 is suggested at MMS 6-NE.			
			These results imply that MMS 6-NE and MMS 6-main could have different dust grain properties. 
			In the map, unreasonably small $\beta$ values, ${\beta}<$0, are shown at the southern part of MMS 6-main, which may be due to the 
			contamination from an extended component. Note that there is extended emission in the 3.3 mm continuum image, which presumably 
			comes from ambient gas along the same line of sight, but no extended emission was detected at 0.9 mm (see Figure 1). 
			
			From the multi-wavelength data, it is suggested that both MMS 6-main and MMS 6-NE are still deeply embedded in the molecular cloud. 
			Nevertheless, the derived value for $\beta$ is low.  
			Draine et al. (2006) shows that the $\beta$ value is highly dependent on grain size; i.e., large grain size gives a smaller $\beta$ value. 
			The observed low-$\beta$ could indicate faster grain growth in the densest region of the molecular core. 
			Actually, recent interferometric observations have also reported $\beta{\sim}$0.5--1 toward Class 0 type stars such as L1448 IRS2, L1448 IRS 3, 
			and L1157 (Kwon et al. 2008), and those results also suggest the faster grain growth. 
						
			In addition to the grain growth, an unresolved optically thick component would also affect the $\beta$ value. 
			The average flux density was derived as 3.4 Jy at 0.9 mm. This corresponds to an averaged brightness temperature of 42 K. 
			Similarly, brightness temperatures were derived from other three-bands data as 12.7 to 15.4 K. Since the opacity, 
			the dust temperature and the observed brightness temperature have a relation of ${T_b}= {T_d} (1-e^{-{\tau_{\nu}}})$, 
			if dust temperature is assumed at the typical temperature of the OMC filaments, $T_d{\sim}$20 K (e.g., Cesaroni \& Wilson 1994), 
			the optical depth of 
			the continuum structure is estimated to be 0.6--2.1, which is partially optically thick. 
			Hence, one possible interpretation to explain the low-$\beta$ is that the larger outer envelope is likely more optically 
			thin and have a higher value of $\beta$, while the inner unresolved component may be more optically thick resulting 
			in a smaller estimated $\beta$ value. Moreover, higher opacity, due to condensed asymmetrical structures such as circumstellar 
			disks inside the cores, can also produce a smaller $\beta$ than the actual $\beta$ as determined by the dust grain 
			properties (e.g., Nakamoto \& Kikuchi 1999; Nakazato, Nakamoto, \& Umemura 2003).

			\subsection{Density and Gravitational Stability}
			In the previous studies, several other compact continuum sources (i.e., 500--2800 AU) were also detected in the 3.3 mm continuum 
			emission toward other OMC-2/3 sources with the NMA (Takahashi et al. 2006, Takahashi et al. 2008a, Shimajiri et al. 2008). 
			In order to compare physical properties of these dust continuum sources with MMS 6-main, the average molecular hydrogen 
			volume density was estimated using the equation of $n=M_{H_2}/(\frac{4}{3}{\pi}R^3{\mu}m_H)$ with the assumption of 
			a spherically symmetrical structure.  
			Here, $M_{H_2}$ is the mass derived from the dust continuum emission, $\mu$ and $m_H$ are the 
			mean molecular weight of 2.33 and hydrogen mass, respectively. Adopting $M_{H2}$=3.0 $M_{\odot}$ and $R{\sim}510$ AU, 
			which were derived in Section 3.2, the mean volume density of MMS 6-main is estimated to be 6.5$\times$10$^9$ cm$^{-3}$. 
			
			Takahashi et al. (2006) have investigated another protostellar core, MMS 7. 
			The 3.3 mm continuum emission was detected in the MMS 7 region with a mass and radius of 0.36--0.72 $M_{\odot}$ and $R{\sim}$500 AU, 
			respectively, and hence the average volume density of H$_2$ was estimated to be 
			(2.6--5.2)$\times$10$^7$ cm$^{-3}$. In addition to MMS 7, the typical hydrogen volume density toward the other dust 
			continuum sources in the OMC-2/3 region was estimated to be several$\times$10$^{(6-7)}$ cm$^{-3}$ from the 3.3 mm 
			continuum emission measured with the NMA (Takahashi et al. 2008a, Shimajiri et al. 2008). 
			While the angular resolutions of most of the previous 3.3 mm continuum observations are somewhat lower than the SMA 
			angular resolution, it is clear that the average hydrogen volume density toward MMS 6-main is roughly one order of 
			magnitude higher than the value derived for MMS 7 and the other OMC-2/3 continuum sources. 
			
			We also compared the MMS 6-main continuum source with the surrounding envelope properties. 
			The average number density of the envelope gas detected in the H$^{13}$CO$^{+}$ emission was 
			estimated to be (1.6--5.0)${\times}$10$^7$ cm$^{-3}$ with a size scale of 5000 AU.			 
			This suggests that the detected compact continuum emission has approximately one order of magnitude smaller size and three  
			order of magnitude higher volume-density as compared with the surrounding envelope structure detected in the H$^{13}$CO$^{+}$ emission. 
			The compact core detected in the interferometric continuum observations is brighter than surrounding parental core because of the high density 
			region towards the center.
			
			Several of the intermediate- to high-mass protostellar core candidates are located in 
			the OMC-1 and OMC-1 south (OMC-1S) regions. In those regions, several bright continuum sources have previously been detected 
			with the high angular resolution of the SMA (Beuther et al. 2004, Zapata et al. 2005). Measured 0.9 mm total fluxes in the OMC-1 
			continuum sources (i.e., source I, source n, SMA 1, and Hot Core) by Beuther et al. (2004) have been reported to be 0.32 to 1.9 Jy. 
			Measured 0.9 mm total fluxes of the strong continuum sources in the OMC-1S region range from approximately 0.3 to 0.8 Jy 
			(L. A. Zapata, private communication based on Zapata et al. 2005). Hence, the comparison of the fluxes between MMS 6-main 
			($F_{\rm{total}}$=3.4 Jy at 0.9 mm) and those continuum sources in the OMC-1 and OMC-1S regions 
			suggests that MMS 6-main is one of the brightest sources in the entire OMC filament in terms of the total flux. 
			
			The next interesting question is whether MMS 6-main is gravitationally bound or not. To discuss this issue, we 
			compared the Virial mass with the mass derived from the interferometric dust-continuum observations. 
			The Virial mass is derived from the equation of 
			$M_{\rm{vir}}{\sim}5RC_{\rm{eff}}^2/G$. Here, the effective sound speed of the gas is derived by 
			$C^2_{\rm{eff}}={\Delta}V^2_{\rm{obs}}/8 {\ln}2$, where $G$ is gravitational constant, and $R$ is radius of 
			the envelope gas derived from the 
			H$^{13}$CO$^{+}$ observations. 
			Using ${\Delta}V_{\rm{obs}}{\sim}$1.0 km s$^{-1}$ which was measured from the H$^{13}$CO$^{+}$(1--0) 
			line profile at MMS 6-main peak with a beam size of 4$''$.6$\times$3$''$.2 (1800$\times$1300 AU), 
			the virial mass was estimated to be 2.3 $M_{\odot}$.
			On the other hand, the H$_2$ mass derived from the 0.9 mm dust continuum emission is 3.0 $M_{\odot}$ within 
			a size scale of $D{\sim}$510 AU. 
			These results show that the H$_2$ mass derived from the 0.9 mm dust continuum emission is higher than the mass 
			derived from the Virial mass estimate over a larger scale, suggesting that MMS 6-main is probably in the gravitationally bound state.  		 
												
			In summary, MMS 6 is one of the brightest submilimeter continuum sources in the entire OMC region. The much higher hydrogen 
			volume density, 6.5${\times}10^9$ cm$^{-3}$, of MMS 6-main is consistent with its gravitationally bound nature. 
			Its density is more than four to five orders of magnitude larger than the average hydrogen volume 
			density of ``typical star-forming dense cores'', which is $\sim$10$^{4-5}$ cm$^{-3}$.

		\subsection{Internal Structure}
			We probe the internal structure of the MMS 6-main core in order to improve our understanding of the initial conditions 
			of the star-forming core. In this  section, we discuss the density distribution of the continuum source, MMS 6-main, 
			using the visibility amplitude plot. 
			The calibrated visibility data are given by the Fourier transform of the intensity distribution 
			of the observed target. In the Fourier domain analysis, a model of the core density structure can be 
			directly compared with the interferometric data without the non-linear image deconvolution process and the missing flux problem. 
			
			For optically thin dust emission, the observed intensity at impact parameter $b$ is given by (e.g., Harvey et al. 2003$a$),
			
				\begin{equation}
						I_{\nu}(b)=2{\int}_{b}^{R_{\rm{out}}} B_{\nu} [T_{d}(r)] {\kappa}_{\nu}(r) {\rho}(r) \frac{r}{(r^2-b^2)^{0.5}} dr, 
				\end{equation}
			
			where $R_{\rm{out}}$ is the outer radius of the system, $T_{d}(r)$ is the dust temperature, ${\rho}(r)$ is 
			the density structure of the system, 
			${\kappa}_{\nu}$ is the dust opacity (mass absorption coefficient), and $B_{\nu}$ is the Planck function. 
			Here, we assume a spherically symmetric spatial structure. 
			If the density and temperature follow simple radial power-laws, ${\rho}(r){\propto}{r^{-p}}$ and $T(r){\propto}{r^{-q}}$, 
			and the opacity follows, ${{\kappa}_{\nu}}{\propto}{\nu}^{\beta}$, 
			the intensity has a simple power-law form, $I({\nu}, b){\propto}{\nu}^{2+{\beta}}b^{-(p+q-1)}$. 
			Since the attenuation resulting from the primary beam antenna pattern of the interferometer is negligible, 
			the visibility distribution is also a power-law (i.e., Fourier transform of the intensity), $V({\nu}, u){\propto}{\nu}^{2+{\beta}}u^{(p+q-3)}$.
			For reasons of expediency, hereafter, we define ${\gamma}=p+q-3$.
					
			Here, the temperature distribution is also important for extracting the density structure since the temperature and density profiles 
			are coupled and observed as a single intensity profile. 
			In the MMS 6-main case, the largest uncertainty comes from the unknown presence of  a central protostar. 
			We consider two kinds of temperature distribution: (i) {\textbf {isothermal temperature distribution}}: 
			No heating source at MMS 6-main since no conclusive evidence for the central heating source of MMS 6-main was 
			obtained from our observations and  (ii) {\textbf {power-law temperature distribution}}: Assumption  
			that the dust is optically thin in the envelope and the heating source (i.e., protostar) is located at the center. 
			 For the latter case where the optically thin dust envelope is heated by a central source, the dust temperature distribution is often expected 
			to follow a power-law index determined by the radius  and emissivity such as $T_d(r){\propto}L^{q/2}r^{-q}$, where $q=2/(4+{\beta})$, 
			where $L$ is the luminosity from the central star (Doty \& Leung 1994). 
			The emissivity value derived in Section 3.3, $\beta$=0.93, for MMS 6-main was adopted, and $q$ was estimated to be 0.41. 
			
			Figure 7 shows the 3.3 mm and 0.9 mm binned visibility amplitudes as a function of the projected $uv$-distance. 
			In order to discuss the density structure of MMS 6-main dusty core, two simple geometrical models, (i)Gaussian shaped density model and 
			(ii)power-law density model, were compared with the observed visibility amplitudes with a baseline range up to 80 k$\lambda$ 
			(corresponds to the core radius of $\sim$450 AU).
			
			The 3.3 mm extended component is fitted fairly well by a 2D-Gaussian model with the size of 4$''$.0$\times$2$''$.8 (1600$\times$1100 AU; 
			denoted by a solid curve in Figure 7$a$). 
			On the other hand, the power-law model applied to the entire $uv$-range (denoted by a dashed curve) is definitely a poor fit. 
			This can be seen to be due to the behaviors at the shortest baselines corresponding  to the larger scale structures. 
			At 3.3 mm, the emission does not rise sharply at the short $uv$-spacings as would be expected for a power-law distribution, consistent with a 
			definite outer radius to the emission structures. 
			A better power-law fit is derived for the 3.3 mm visibility data restricted to larger baselines, ${\geq}12$ k$\lambda$ (see a doted curve). 
			This can be explained by the absence  of short spacing information to constrain the extended component. 
			As in the case for the 3.3 mm data, the extended structure in the 0.9 mm data is fitted well with either the 2D-Gaussian model 
			(with a size of $2''.3{\times}2''.0$, corresponds to 920$\times$800 AU) or the power-law density distribution up to the uv-distance 
			of 18 k$\lambda$. These results suggest that the extended halo around MMS 6-main has a definite outer radius of 
			2000-3000 AU (corresponds to 12--18 k$\lambda$). 		
			
			The 3.3 mm extended component is fitted with a power-law index of $\gamma$=-0.55. Since $p=3-q+{\gamma}$, we estimate $p$=2.45 for 
			the isothermal case (i.e, $q$=0) and $p$=2.00 for the power-law temperature case (i.e., $q$=0.41).  
			The estimated density profile from the interferometric data, with $p$=2.00--2.45, is roughly consistent with the large-scale (10000 AU) 
			density structure, with $p{\sim}2$, as derived from  previous single-dish 850 $\mu$m observations by JCMT/SCUBA (J$\o$rgensen et al. 2006). 
			The observed density power-law index toward MMS 6-main, therefore, roughly agrees with a quasi-static star-forming scenario 
			with a singular isothermal sphere (SIS), which has a density distribution of $\rho(r){\propto}r^{-2}$ (Larson 1969; Penston 1969), 
			or even has a steeper power-law index than the isothermal case. 
			There is indication that a steeper density distribution is established as the spatial size-scale decreases toward 450 AU. 
 			In the visibility fitting, the 0.9 mm data show a very similar amplitude power-law index ($\gamma$=0.61) and hence similar 
			density power-law index ($p$=1.98--2.39). 			
			
			One feature to note is that no flattened density structure was detected at the core center, which is usually expected toward prestellar cores
			\footnote{We have 7.3 mm data, which have $uv$-ranges from 4.5 to 477 k$\lambda$. However, there are only sufficient sensitivities 
			for baselines shorter than $\sim$100 k$\lambda$. Hence the 7.3 mm data did not add new information on the compact component, 
			as suggested in the 0.9 mm and 3.3 mm data} (e.g., Ward-Thompson et al. 1994). Non-detection of the flattened density structure implies 
			two possibilities; (i) current angular resolution is still too coarse, as a centrally flattened structure might exist  with a size scale much less
			 than $R{\sim}450$ AU (corresponds to 80 $k{\lambda}$), or (ii) MMS 6-main already has a central protostar, 
			but the dynamically collapsing area, which has a density distribution of ${\rho}(r){\propto}r^{-1.5}$ such as proposed by Shu et al. (1977), 
			is still compact (i.e., $R<$450 AU). In either case, a constant power-law index suggests that MMS 6-main is in either (i) the latest stage of 
			prestellar core or (ii) the earliest stage of protostellar core.

			Observationally, the density profile maintains the same power-law index up to the $uv$-distance of 80 k$\lambda$, 
			which corresponds to  the linear size of $R{\sim}$450 AU. This implies that the upper limit to the size of the infalling envelope is 
			$R{\sim}$450 AU. $C_{\rm{eff}}{\sim}$0.28 km s$^{-1}$ as derived from the assumption of $T{\sim}20$ K, 
			then implies an upper limit to the age of the protostar of ${\tau_{{core}}}{\leq}R/C_{eff}{\sim}$7.6$\times$10$^{3}$ yr. 
			On the other hand, the free-fall time of MMS 6-main as derived from the average density; ${\tau}_{ff}=1/{\sqrt{G{\rho}}}{\sim}
			2.12{\times}10^{5}(n_{\rm{H_2}}/10^5 cm^{-3})^{-1/2}{\sim}$830 yr. 
			Both timescales are much shorter than the typical lifetime of the Class 0/I protostars, which is ${\sim}10^{(4-5)}$ yr 
			(Beichman et al. 1986; Cohen, Emerson, \& Beichman 1989; Kenyon et al. 1990; Beichman, Boulanger, \& Moshir 1992).

			In addition to the extended envelope, a compact source, which appears as a remnant unresolved component, was suggested 
			from the Gaussian fitting of the 3.3 mm and 0.9 mm data (denoted by Figure 7$a,b$ dashed line). 
			Although the observed $uv$-coverage is not enough to clearly show this component, the constant or slowly decreasing 
			visibility amplitudes at long baselines with $uv$-distances of more than 60 $k{\lambda}$ (corresponds to a linear size scale 
			of $D{\sim}$1200 AU) suggest the presence of this compact source. In contrast, power law-density fittings do not require an 
			unresolved component. Longer baseline data with a $uv$-distance more than 80 k$\lambda$ are 
			necessary to judge whether the unresolved component exists or not. 
			From the Gaussian fitting, the best-fitted offset flux is derived as 56 mJy at 3.3 mm and 1.6 Jy at 0.9 mm, 
			which suggests an upper limit of H$_2$ mass of 0.2--1.0 $M_{\odot}$ under the assumption of $T_d$=20--100 K.

			\section{Conclusion and Future Prospects: 
			MMS 6-main is an Extremely Young Stage of the Protostellar Core?}
			The millimeter and submillimeter multi-wavelength continuum observations with the CO (1--0) and H$^{13}$CO$^{+}$(1--0)  
			line emissions have been performed with the NMA, SMA, and VLA. The main results are summarized as follows; 
	
			\begin{itemize}		
			\item	A compact continuum source, MMS 6-main was detected with an apparent H$_2$ mass of 
			3.0 $M_{\odot}$ and a typical size of 510 AU.  
			Despite its compact appearance, no clear evidence of a radio jet or a CO molecular outflow were detected at the position of MMS 6-main.
			Moreover, from the multi-wavelength continuum images, no infrared source is seen at a wavelength shorter than 8 $\mu$m 
			at the peak position of MMS 6-main. 
			The derived H$_2$ column density, 2.6$\times$10${^{25}}$ cm$^{-2}$, corresponds to the visual extinction of $A_v{\sim}$15000 mag 
			and $A_{8{\mu}m}$=790 mag. The estimated $A_{8{\mu}m}$ toward MMS 6-main is much higher than those derived toward 
			other YSOs in the OMC filament, 
			$\sim$10 mag. Therefore, it is possible that the central protostar 
			associated with MMS 6-main cannot be detected because of the extremely high extinction. 
			
			\item	At least two orders of magnitude higher volume density, ${\sim}6.5{\times}10^9$ cm$^{-3}$, as compared with the 
			other OMC-2/3 continuum sources, was suggested toward MMS 6-main.
			The internal structure of MMS 6-main roughly agrees with an isothermal dusty core with a power-law index of $p{\sim}$2. 
			We note that the density structure with $p{\sim}$2 continues up to the $uv$-distance of 80 k$\lambda$, 
			which corresponds to the linear size of $R{\sim}$450 AU. 
			Assuming that the upper limit of the infalling radius is $R{\sim}$450 AU, the upper limit of the protostellar age is estimated to be 
			$\leq$7.6$\times$10$^{3}$ yr, while the free-fall time scale of the core is estimated to be ${\sim}830$ yr from the 
			average dust density. Both timescales are roughly consistent with each other, and the estimated timescale is extremely short as compared 
			with the typical time scale of the Class 0/I protostars which is ${\sim}10^{(4-5)}$ yr. This suggests that MMS 6-main is probably in the 
			pre-stellar phase or a transition phase from the pre-stellar to the proto-stellar evolutionary stage. 
			An unresolved component was suggested from the Gaussian fitting of the 3.3 mm and 0.9 mm data with an upper limit to the mass 
			of 0.2--1.0 $M_{\odot}$. The alternative power-law density fitting would not require such a compact component. 
			Longer baseline data with a $uv$-distance more than 80 k$\lambda$ are necessary to constrain the inner most structure. 
			
			\end{itemize}

			Due to the lack of even higher angular resolution, sensitivity, and an appropriate molecular tracer, 
			we could not directly measure the physical properties of the 100 AU scale circumstellar disk or the possible central protostar. 
			SMA and VLA observations with higher angular resolutions and also future ALMA observations will reveal the internal structure of 
			MMS 6-main with a size-scale down to several AU. 
			High angular resolution data of H$_2$O masers could be useful for searching for the very beginning 
			of the mass ejection phenomena (i.e., the micro-jet) from the central protostar such as for the S106 FIR case (Furuya et al. 1999). 
			Moreover, polarization observations with higher spatial resolutions 
			can be useful for elucidating the magnetic field structure at this early evolutionary stage. 
			In addition to the observational studies, comparison with models, which 
			predict the very early stages of star-forming cores such as Bonner-Ebert spheres or adiabatic "First-Cores" 
			(e.g., Larson 1969; Masunaga et al. 1998; Saigo 2006, 2008) will also be important.  

\acknowledgments
We acknowledge the staff at the Nobeyama Radio Observatory, the Submillimeter Array, and the National Radio Astronomy Observatory, 
for assistance with operations and data reductions. 
The authors thank J. Lim for his early contributions to the VLA observations, and C.-H. Yan for his support with 
the infrared data analysis. S. T. is deeply grateful to L. A. Zapata for the information provided on continuum source data in the OMC 1-South region. 
S. T. acknowledges Y. N. Su, B. A. Whitney, K. Asada, N. Ohashi, W. P. Chen and J. H.S. Lin for helpful discussions. 
We also thank an anonymous referee whose suggestions improved this manuscript. 
This publication used archival data from the Spitzer Space Telescope and the Very Large Array. 
A part of this study was financially supported by Grant in-Aid for Scientific Research 18204017 (or Global COE Program 
``the Physical Science Frontier'' MEXT, Japan). 
S. T. was financially supported by the Japan Society for the promotion of Science (JSPS) for Young Scientists, and 
is financially supported by a postdoctoral fellowship at the Institute of Astronomy and Astrophysics, Academia Sinica, Taiwan.

\clearpage

\begin{figure}
\rotate
\epsscale{1.0}
\plotone{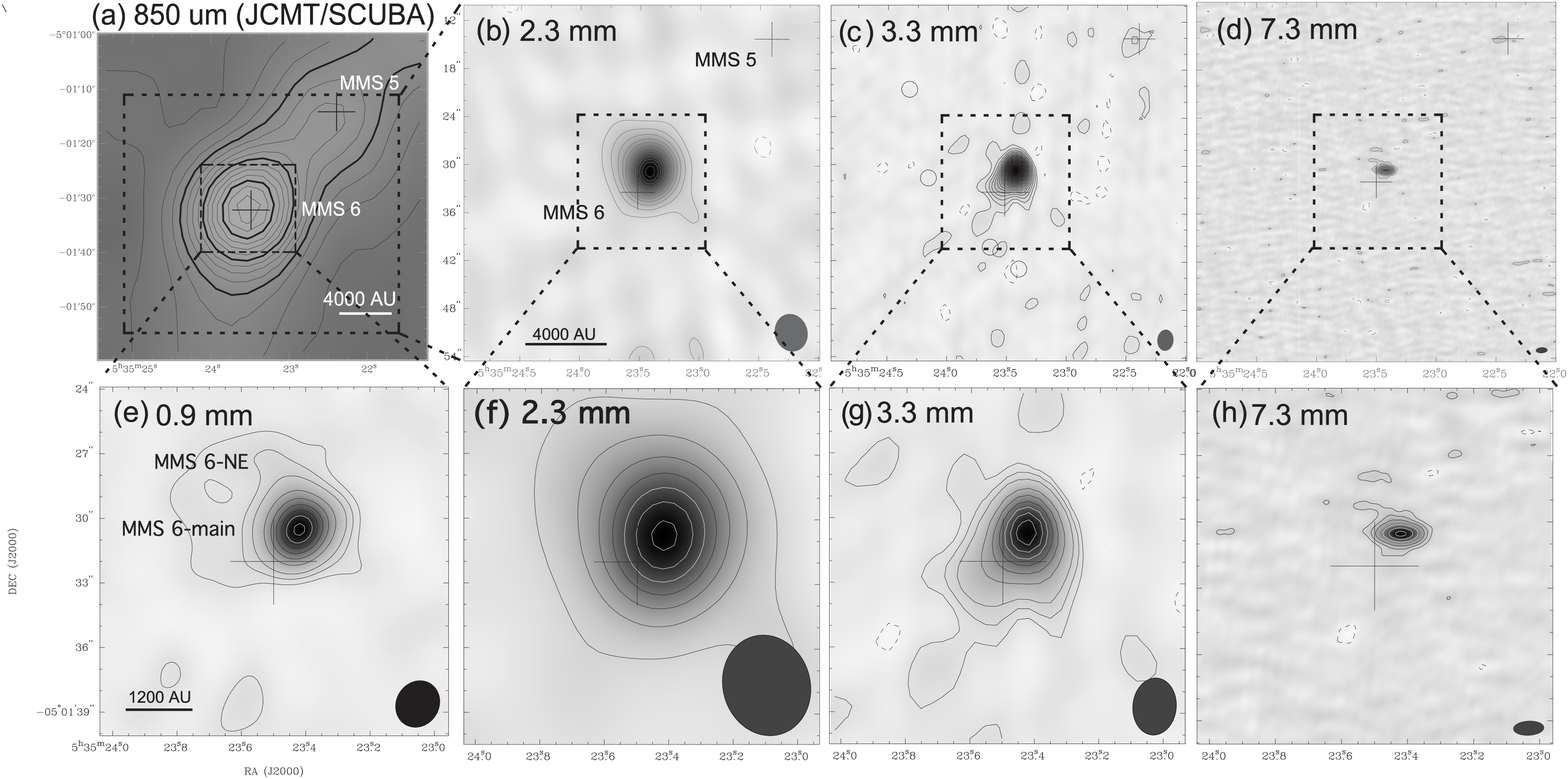}
\caption{(a); 850 $\mu$m continuum image taken with SCUBA/JCMT (Johnstone et al. 1999). 
The contour intervals are 0.5 Jy beam$^{-1}$ beginning at 0.5 Jy beam $^{-1}$. 
(b), (c), and (d); Map of 3.3, 2.3, and 7.3 mm continuum images in the MMS 6/5 regions (large scale). 
The contour levels start at 3$\sigma$ level, with an interval of 3$\sigma$. 
The maps were made using all $uv$-coverage with robust weighting. 
(e), (f), (g), and (h); Focused map of 0.9, 3.3, 2.3, and 7.3 mm continuum images around MMS 6-main. 
The contour levels are -3, 3, 6, 10, 20, 30$\sigma$, ..... 
Negative contours are shown by dashed lines. The crosses show the peak positions of the 1.3 mm continuum 
image identified by IRAM 30 m (Chini et al. 1997). Filled ellipses at the bottom right corners show the synthesized beams of each map. 
The detailed parameters of these maps are presented in Table 2. 
\label{f1}}
\end{figure}
\clearpage

\begin{figure}
\rotate
\epsscale{1.0}
\plotone{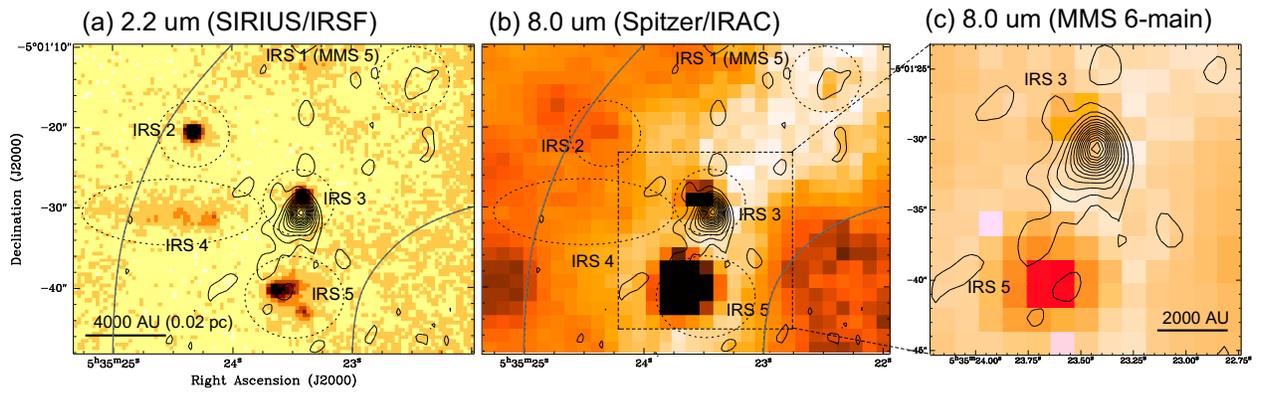}
\caption{(a) 2.2 $\mu$m-band and (b) 8.0 $\mu$m images overlaid with the 3.3 mm continuum maps made with the NMA (contours start at 3-$\sigma$ level with 
an interval of 6 $\sigma$). (c) Focused image of Figure (b). Identified infrared sources are denoted by the dashed lines. 
Thin solid lines in the images show the boundary area for the infrared source identifications, which is roughly consistent with the distribution of the OMC filamentary structure. 
\label{f2}}
\end{figure}
\clearpage

\begin{figure}
\rotate
\epsscale{0.70}
\plotone{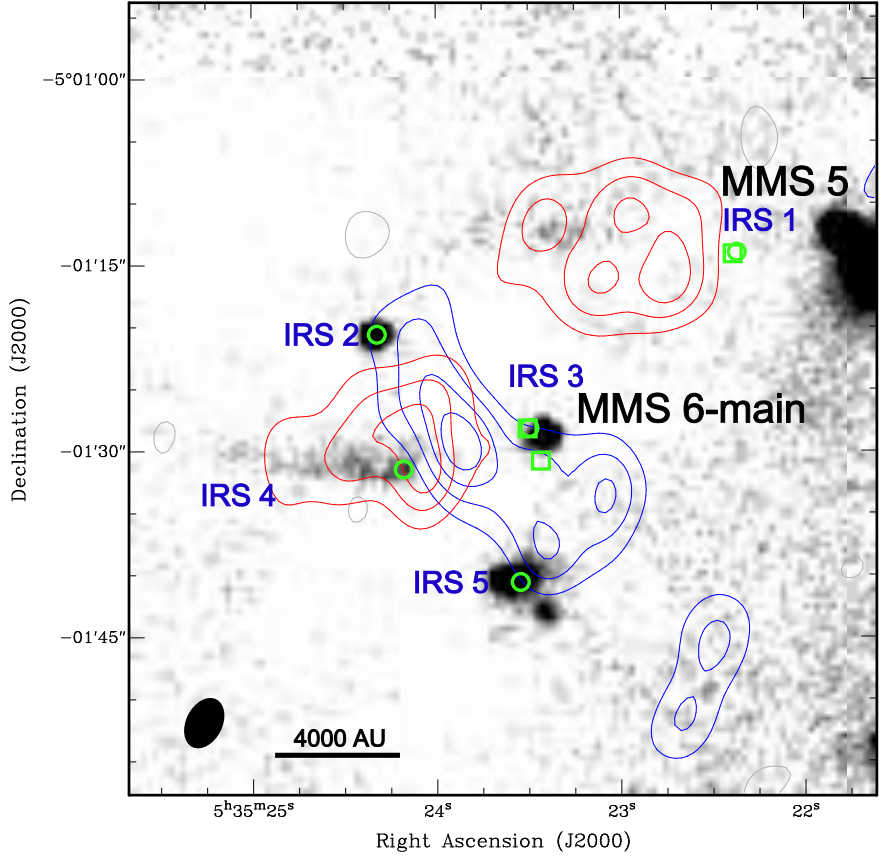}
\caption{Total intensity map of the CO (1--0) emission in contours, made with the NMA, superposed on the $Ks$-image in gray scale taken with SIRIUS/IRSF.
The velocity ranges of the blueshifted and redshifted emission are $V_{\rm{LSR}}$ = 8.3 to 10.3 km s$^{-1}$ and 
$V_{\rm{LSR}}$ = 12.5 to 14.5 km s$^{-1}$, respectively.
A filled ellipse at the bottom left corner shows the synthesized beam of the CO (1--0) emission.
Open squares show peak positions of MMS 5, MMS 6-main and MMS 6-NE  as detected by the interferometric data.  
Open circles show positions of identified infrared sources (i.e., IRS 1 to IRS 5) measured at Spitzer/IRAC bands. 
\label{f3}}
\end{figure}
\clearpage

\begin{figure}
\rotate
\epsscale{1.0}
\plotone{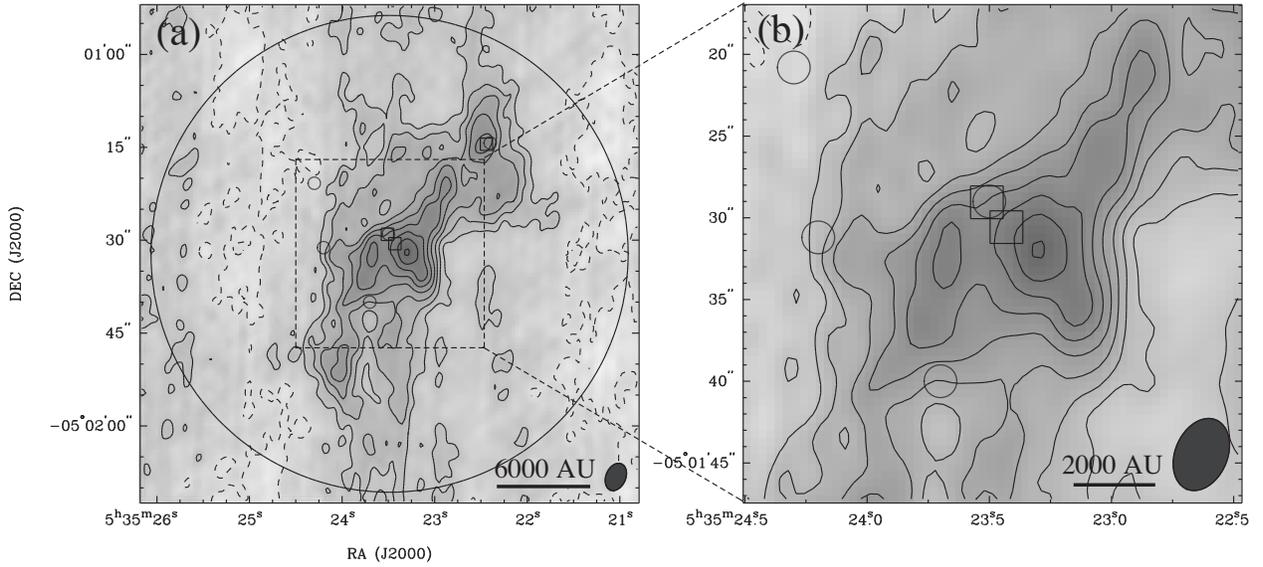}
\caption{(a) Total intensity map of the H$^{13}$CO$^{+}$ (1--0) emission integrated from $V_{\rm{LSR}}$ = 10.8 to 12.1 km s$^{-1}$ (Large scale).
The contour levels start at 2$\sigma$ level, with an interval of 2$\sigma$ (1-$\sigma$= 0.24 Jy beam$^{-1}$ km s$^{-1}$). Negative contours are shown by dashed lines. 
Open-squares and open-circles show positions of milllimeter/submillimeter sources; MMS 5/6; and infrared sources (i.e., IRS 1, 2, 3, 4, and 5), respectively. 
Filled ellipse at the bottom left corner shows the synthesized beam. Large circle on the map denotes the FWHM size of the primary beam. 
Figure (b) shows focused map of Figure (a).
\label{f4}}
\end{figure}
\clearpage

\begin{figure}
\rotate
\epsscale{0.8}
\plotone{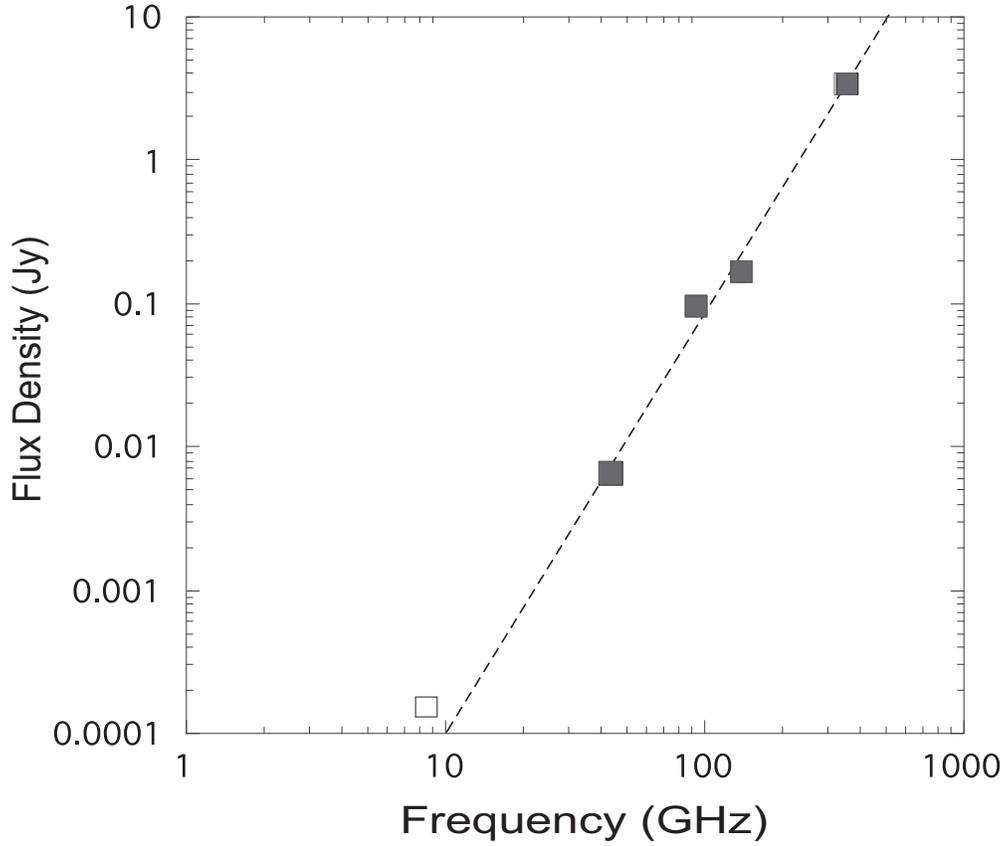}
\caption{Flux density of MMS 6-main as a function of the frequency. 
Black (filled and open) squares show total flux densities of each frequency measured from the images. 
The measured values are presented in Table 4. 8.4 GHz flux was referred to Reipurth et al. (1999; taken with VLA D-array observations) as
an upper limit value. Each flux has an error bar within each square. The best fitting result is denoted by dashed line. 
\label{f1}}
\end{figure}
\clearpage

\begin{figure}
\rotate
\epsscale{0.7}
\plotone{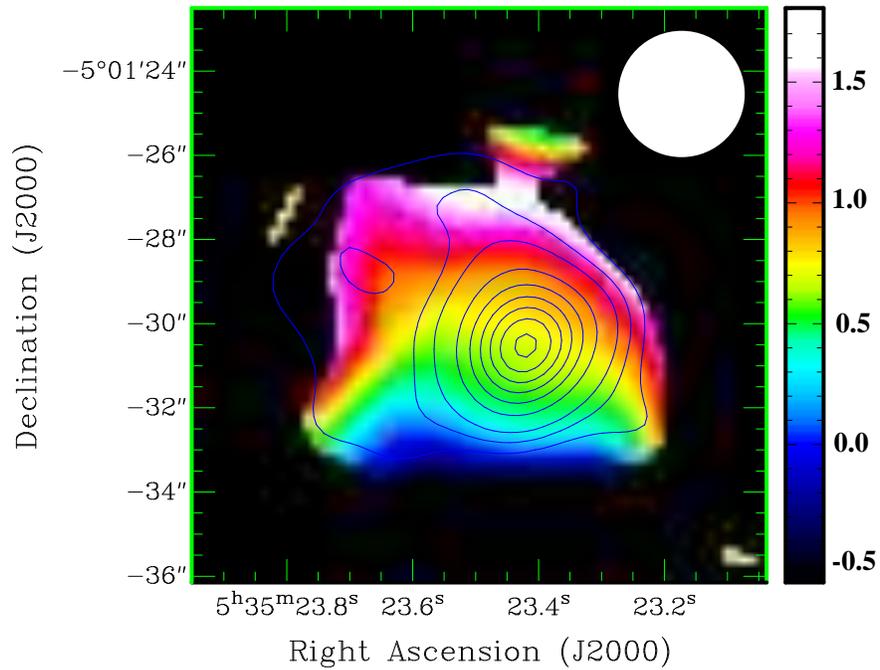}
\caption{Dust emissivity index ($\beta$) maps of the MMS 6 region (color) derived from the 3.3 mm and 0.9 mm continuum maps overlaid with the 0.9 mm continuum image (contours). 
Both 3.3 mm and 0.9 mm continuum maps were convolved with the 3$''$ beam size (see upper right corner), and values for 
$\beta$ were calculated using the equation (1). 
\label{f1}}
\end{figure}
\clearpage

\begin{figure}
\rotate
\epsscale{0.7}
\plotone{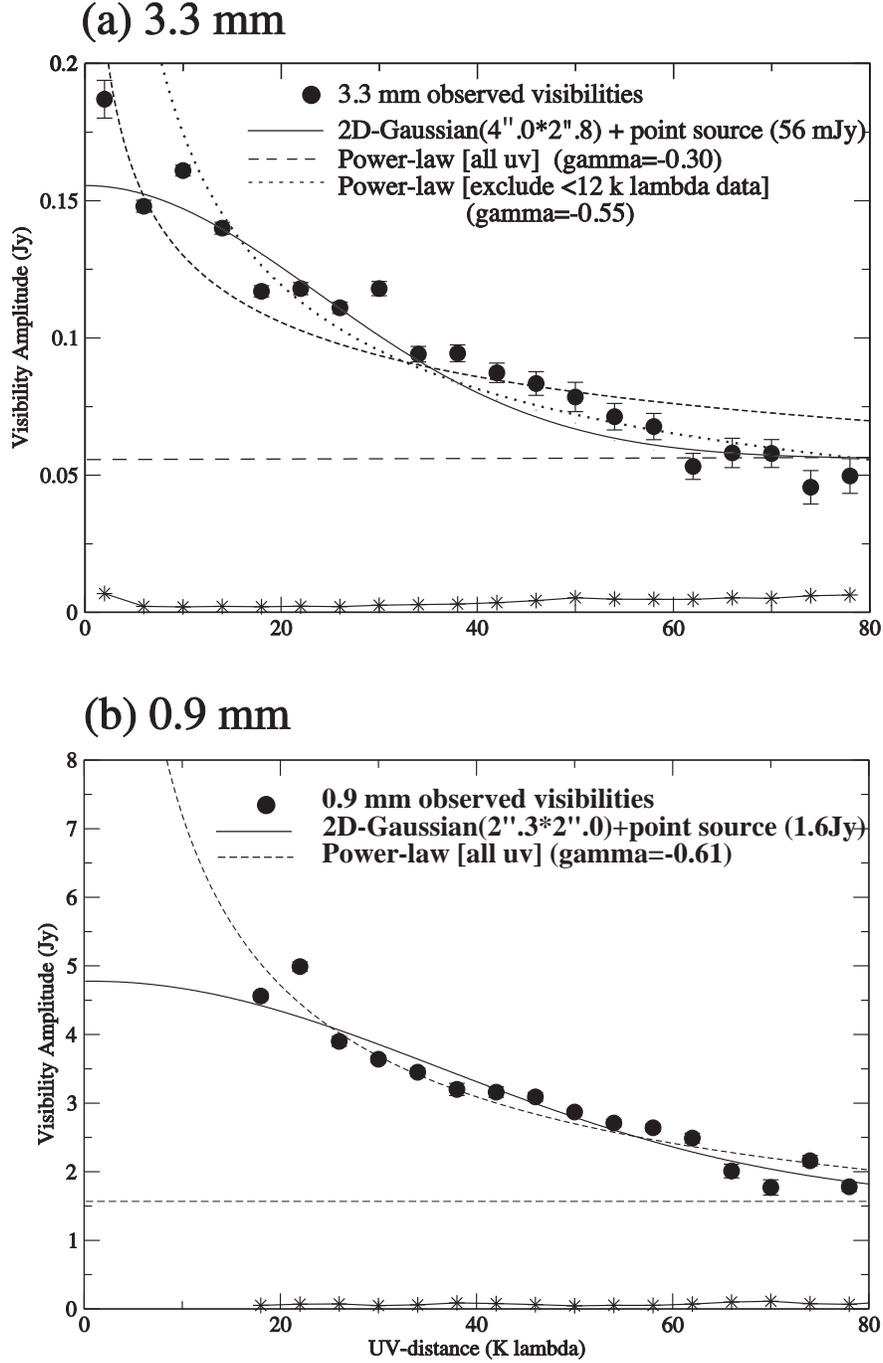}
\caption{4 $k{\lambda}$ binned visibility amplitudes as a function of $uv$-distance for $(a):$ 3.3 mm and $(b):$ 0.9 mm continuum data, 
together with the best-fit models. Gaussian+point sources fitting results are denoted by solid lines. Power-law fitting results are denoted by dashed lines. 
For Figure $(a)$, two power-law fits; (i) the best power-law fit derived from all the visibility data was denoted by the dashed line, 
and (ii) the best power-law fit derived from the visibility data ranging from 12 k$\lambda$ to 80 k$\lambda$ (same as the SMA $uv$-coverage) 
was denoted by the dotted line. Horizontal dashed lines in the Figures show the best-fit offset values for the possible presence of an unresolved component. 
1$\sigma$ statistical errors were denoted by asterisks with solid line.
\label{f9}}
\end{figure}
\clearpage

\begin{deluxetable}{lccccc}
\rotate
\tabletypesize{\scriptsize}
\tablecaption{Observational Parameters of Continuum Emissions\label{tbl2}}
\tablewidth{0pt}
\tablehead{
\colhead{Parameter} & \colhead{0.9 mm} & \colhead{2.3 mm} & \colhead{3.3 mm} & \colhead{7.3 mm} & \colhead{3.6 cm}\\
}
\startdata
Telescope			                                              &  SMA & NMA & NMA & VLA  &VLA \\
Configurations\tablenotemark{a}	  	                  &  compact & D & AB, C, and D  & C and DnC & DnC \\
Primary beam HPBW (arcsec)	  			           &  36  & 46  &  77  &  60  & 320 \\
Synthesized Beam HPBW (arcsec)	  	           &  2.2$\times$2.0 (-32$^{\circ}$) & 4.7$\times$4.0 (16$^{\circ}$) 
& 2.6$\times$2.0 (-4.8$^{\circ}$) & 1.4$\times$0.6 (-87$^{\circ}$) & 10$\times$3.0 (67$^{\circ}$)\\ 
Projected base line range (k$\lambda$)       &  13--88    & 4 -- 38  &  3 -- 115  &  4.5 -- 477 & 8.7 -- 55 \\
Maximum detectable structure (arcsec)\tablenotemark{b}        &  13  &  35  &  56  &  43  & 180\\
Bandwidth (GHz)	                                             &  4  &  2  &  2  &  0.1 & 0.1\\ 
Gain calibrator                      &  0530+135  & 0530+135  &  0530+135 & 05416-05418 & 05416-05418 \\
Flux of the gain calibrator (Jy)                        &  1.7  &  4.0  &  2.5  &  0.7  & 1.4\\
Bandpass calibrator	                                        &  3C273 and 3C454.3  &  0530+135  &  3C454.3 & --- & ---\\
Flux calibrator                                                  &  Titan  &  Uranus  &  Uranus  &  3C286 and 3C48 & 3C48  \\  
System temperature in DSB (K)\tablenotemark{c}                      &  200 -- 400  & 300 -- 800  & 240 -- 600 & ---\tablenotemark{d} & ---\tablenotemark{d} \\ 
RMS noise level (mJy beam$^{-1}$)                 &  43  &  6.5  &  1.0  &  0.1  & 0.05 -- 0.06 \\
\enddata
\tablenotetext{a}{D and AB are most compact and sparse configurations, respectively.}
\tablenotetext{b}{Our millimeter and submillimeter observations were insensitive to more extended emission than this size scale structure at the 10 \% 
level (Wilner \& Welch 1994). The maximum detectable structure for the VLA observations were referred from the VLA home page
(http://www.vla.nrao.edu/astro/guides/vlas/current/node10.html).}
\tablenotetext{c}{System noise temperature of the SIS receiver was measured at the source elevation angle.}
\tablenotetext{d}{In stead of the system noise temperature information, VLA observation log including weather and system conditions is available 
on the VLA home page (http://www.vla.nrao.edu/operators/logs/2008/06/2008-06-01\_1758\_AT359.pdf)}
\end{deluxetable}

\begin{deluxetable}{lcc}
\tabletypesize{\scriptsize}
\tablecaption{Observational Parameters of Line Emissions observed with the NMA\label{tbl1}}
\tablewidth{0pt}
\tablehead{
\colhead{Parameter} & \colhead{H$^{13}$CO$^{+}$ (1--0)} & \colhead{$^{12}$CO (1--0)} \\}
\startdata
Configuration\tablenotemark{a}			&	C and D                                           &  D   	\\
Baseline (k$\lambda$)	  					   &	2.9 to 100 	                                &	7.8 to 30    \\
Primary beam HPBW (arcsec)	  			 &     77 	                                              &  62    		  \\
Synthesized Beam HPBW (arcsec)	  	  &	   4.6$\times$3.2 (-23$^{\circ}$) 	     &  6.1$\times$4.4 (-13$^{\circ}$)  \\
Maximum detectable structure (arcsec)\tablenotemark{b} &   60    &  21  \\
Velocity Resolution (km s$^{-1}$)	     &	  0.22 	                                         &  2.0 \\
Gain calibrator\tablenotemark{c}	     &	 0531+135 	                                   &  0531+135 	 	   \\
Bandpass calibrator	\tablenotemark{d} &	 3C 454.3	                                     &  3C 273						          \\
System temperature in DSB (K) \tablenotemark{e}    &    300 -- 600	            &  200 -- 600						      \\
RMS noise level (Jy beam$^{-1}$)         &    9.2e-02 	                                       &	 2.0e-01 	 \\
\enddata
\tablenotetext{a}{D configuration is the most compact configuration.}
\tablenotetext{b}{Our observations were insensitive to more extended emission than this size scale structure at the 10\% level (Wilner \& Welch 1994).}
\tablenotetext{c}{Gain calibrators were observed every 20 minutes. The flux density of the gain calibrator was determined relative to Uranus.}
\tablenotetext{d}{Bandpass calibrations were achieved by 30 to 40 minutes observations.}
\tablenotetext{e}{System noise temperature of the SIS receiver was measured at the source elevation angle.}
\end{deluxetable}

\begin{deluxetable}{lcccccc}
\tabletypesize{\scriptsize}
\tablecaption{Gaussian image fitting results toward MMS 6-main\label{tbl3}}
\tablewidth{0pt}
\tablehead{
\colhead{Wavelength} & \colhead{R.A. (J2000)} & \colhead{Dec. (J2000)} & \colhead{Total flux} 
& \colhead{Peak flux} & \colhead{Deconvolved size\tablenotemark{a}} & \colhead{P.A.} \\
\colhead{(mm)} & \colhead{} &\colhead{} & \colhead{(mJy)} & \colhead{(mJy beam$^{-1}$)} & \colhead{(arcsec)} & \colhead{(degree)}\\
}
\startdata
\multicolumn{7}{c}{MMS 6-main}  \\
\hline
0.9 mm &  5$^h$35$^m$23$^s$.42  &	-5$^{\circ}$01$'$30$''$.54	   & 4.0e+03      &    3.3e+03    & 2.4$\times$2.1  & -28 \\  
2.3 mm &  5$^h$35$^m$23$^s$.42  &	-5$^{\circ}$01$'$30$''$.61	   & 4.5e+02      &    2.5e+02   &  4.1$\times$3.7  &  172     \\
3.3 mm &  5$^h$35$^m$23$^s$.43  &	-5$^{\circ}$01$'$30$''$.80	   & 1.4e+02      &    8.0e+01   & 2.2$\times$1.8  & -31 \\  
7.3 mm &  5$^h$35$^m$23$^s$.42  &	-5$^{\circ}$01$'$30$''$.52	   & 6.0e+00      &    4.1e+00   &  0.8$\times$0.5  & 67 \\
\hline
\multicolumn{7}{c}{MMS 6-NE}  \\
\hline
0.9 mm\tablenotemark{b} &  5$^h$35$^m$23$^s$.50  &	-5$^{\circ}$01$'$29$''$.58	        &  1.6e+03      &    3.2e+02    & 5.5$\times$3.8  & -15 \\  
\hline
\multicolumn{7}{c}{MMS 5}  \\
\hline
3.3 mm &  5$^h$35$^m$22$^s$.44  &	-5$^{\circ}$01$'$14$''$.50	        &---\tablenotemark{b}    & 1.2e+01    & ---\tablenotemark{b}  & ---\tablenotemark{b} \\  
7.3 mm &  5$^h$35$^m$22$^s$.47  &	-5$^{\circ}$01$'$14$''$.35	        &---\tablenotemark{b}    & 1.4e+00    & ---\tablenotemark{b}  & ---\tablenotemark{b} \\  
\enddata
\tablenotetext{a}{1$''$ corresponds to 400 AU at  the Orion Molecular Cloud.}
\tablenotetext{b}{Deconvolved size could not be derived from the 3.3 mm and 7.7 mm continuum data.}
\end{deluxetable}

\begin{deluxetable}{llll}
\tabletypesize{\scriptsize}
\tablecaption{Identified infrared sources in the MMS 5/6 region\label{}}
\tablewidth{0pt}
\tablehead{
\colhead{Source Name} & \colhead{RA.} & \colhead{Dec.} & \colhead{Note\tablenotemark{a}} \\
\colhead{} & \colhead{(J2000)} & \colhead{(J2000)} & \colhead{} \\
}
\startdata
IRS 1 (MMS 5) & 5 35 22.4 & -5 01 14.3 & Class 0  \\
IRS 2 & 5 35 24.3 & -5 01 20.8 & PMS\\
IRS 3 & 5 35 23.5 & -5 01 29.0 & Class I\\
IRS 4 \tablenotemark{b} & 5 35 24.2 & -5 01 31.2  & Reflection nebula\\
IRS 5 & 5 35 23.7 & -5 01 40.0 & Class I\\
\enddata  
\tablenotetext{a}{Evolutionary status of each source was determined from the Spectral Energy Distribution.}
\tablenotetext{b}{IRS 4 has extended emission, and its position was measured at the brightest part of the near-infrared structure.}
\end{deluxetable}

\begin{deluxetable}{llll}
\tabletypesize{\scriptsize}
\tablecaption{Flux density of MMS 6 main\label{tbl4}}
\tablewidth{0pt}
\tablehead{
\colhead{Wavelength} & \colhead{Flux density} & \colhead{Error} & \colhead{Deconvoled Size (P.A.)} \\
\colhead{($\mu$m)} & \colhead{(mJy)} & \colhead{(mJy)} & \colhead{(arcsec$\times$arcsec),(degree)} \\
}
\startdata
3.6 cm\tablenotemark{a}     &  $<$1.5e-1    &   4.0e-02        &   ---\tablenotemark{d}   \\
7.3 mm    &  6.6e+00 \tablenotemark{b,c}   &   1.9e-01        &    0.8$\times$0.4 (160)  \\
3.3 mm    &  9.6e+01 \tablenotemark{b,c}    &   3.5e+00      &   1.1$\times$1.0 (106)    \\
2.3 mm    &  1.7e+02 \tablenotemark{b,c}    &   2.6e+01      &   ---\tablenotemark{d}    \\
0.9 mm    &  3.4e+03       &   1.0e+02     &  1.0$\times$0.8 (166)    \\
\enddata
\tablenotetext{a}{Referred from Reipurth et al. (1999)}
\tablenotetext{b}{The flux densities were measured using uv-data which are longer than 30 k$\lambda$.}
\tablenotetext{c}{The values are different from the values derived from the gaussian image fitting (Table 3) because we applied 
the gaussian fitting using the different uv-range and also different method.}
\tablenotetext{d}{Deconvolved size could not be derived from the data.}
\end{deluxetable}

\end{document}